\newcommand{\cv}{D_{\delta}}
\documentclass[12pt]{iopart}
\usepackage{graphicx}
\usepackage{float}
\usepackage{color}

\usepackage{iopams}  
\usepackage{cancel}
\begin{document}

\title[Regge-Teitelboim cosmology driven by a curved background]{Regge-Teitelboim cosmology driven by a curved
background}

\author{Giovany Cruz${}^\dag$ \& Efraín Rojas${}^*$}

\address{Facultad de Física, Universidad Veracruzana, Paseo No. 112,
Desarrollo Habitacional Nuevo Xalapa, Xalapa-Enríquez, 91097, Veracruz, México}
\ead{giocruz@uv.mx${}^\dag$, efrojas@uv.mx${}^*$}
\vspace{10pt}
\begin{indented}
\item[]September 26, 2024
\end{indented}

\begin{abstract}
We derive the dynamic equations governing a Regge-Teitelboim (RT) brane that represents our universe considering that its degrees of freedom are the embedding functions, evolving in a five-dimensional
curved spacetime. Within this framework, we investigate the effects of embedding a Friedmann-Robertson-Walker (FRW) metric into a specific curved ambient spacetime. This approach allows us to analyze in detail how the ambient spacetime's curvature influences the brane universe's dynamics. Specifically, we examine the relationship between the Hubble parameter $H$ and the redshift $z$, demonstrating that our results significantly agree with actual observational data. This is contrasted and compared with other approaches.
\end{abstract}

%
%
%
%
%

\section{Introduction}

The use of branes to describe various physical systems has gained significant prominence. In M-theory, they are regarded as fundamental entities \cite{simon2012brane}. On the other side, 
at large scales, the concept of brane plays a central role in the braneworld scenario, which envisions the universe as an extended object embedded in a higher-dimensional spacetime \cite{ida2000brane,maartens2001geometry}. Within this framework, a particularly notable model is that proposed by Tullio Regge and Claudio Teitelboim in the 1970s \cite{regge2016general}. They suggested that our universe can be conceptualized as a brane embedded in a flat Minkowski spacetime, with its degrees of freedom represented by the embedding functions that describe the brane floating in the ambient spacetime. In this framework, Einstein's equations of General Relativity (GR) are not fundamental; they are traded for a compact oriented form, namely, $G^{ab} K_{ab} = 0$. However, all GR solutions are encapsulated within the dynamic equations governing the brane. 
Regarding the geometric requirements for embeddings a fully general four-dimensional metric necessitates considering a Minkowski spacetime of at least ten dimensions, as dictated by the embedding theorems \cite{friedman1961local}. Fortunately, this requirement can be relaxed when the ambient spacetime is not Minkowski. For example, it is proved in \cite{romero1996embedding} that any four-dimensional metric that is a solution to GR can be locally embedded in a five-dimensional Ricci-flat ambient spacetime.

In this work, we address the issue of investigate
how the presence of a curved background spacetime can account
for the acceleration effects of the universe taking $\Lambda$CDM model as a guide to compare with current observations. To achieve this we will consider that the universe is
governed by the RT action and embedded in a dynamic five-dimensional ambient spacetime. The field variables of this effective field theory include the components of the ambient spacetime metric, the matter fields within the
background spacetime, the embedding functions, and the matter fields on the brane. We focus then in investigating the effects that the curvature of the ambient space might have on the dynamical equations of the brane, in the same
spirit of \cite{stern2023dark}. 

This paper is organized as follows. In Section 2 we introduce the notation used throughout the work and discuss key geometric concepts relevant to describing extended objects. We also address the covariant variation of fundamental geometric entities, which help to derive the equations of motion. Section 3 is devoted to explaining the Regge-Teitelboim model within the context of dynamically curved ambient spacetime. We embed an FRW spacetime into a five-dimensional de-Sitter spacetime and compare the results obtained with observational data in Section 4. The final Section presents the conclusions drawn from our findings.

\section{\label{level2}Geometry and notation}

Consider that the world volume describing the evolution of a brane is a four-dimensional manifold, denoted by $m$. This manifold is embedded in a curved five-dimensional ambient spacetime $\mathcal{M}$ with a metric denoted by $G_{\mu\nu}$ where a covariant derivative, $\mathcal{D}_{\mu}$, is compatible with the metric, i.e., 
$\mathcal{D}_{\alpha} G_{\mu\nu} = 0$. The world volume can be described by embeddings $y^{\mu} = X^{\mu}(\xi^a)$, where $y^{\mu}$ are the coordinates of $\mathcal{M}$, $X^{\mu}$ are the embedding functions, and $\xi^a$ are the coordinates on $m$. Tangent vectors to $m$ can be constructed by taking the derivative of the embedding functions with respect to the coordinates $\xi^a$
\begin{equation}
X_a^{\mu}:=\frac{\partial X^{\mu}}{\partial \xi^a}.
\label{comp_tangent_vectors}
\end{equation}
These form a tangent base at each point of the world volume. Note that we are using Latin indices for $m$ and Greek indices for $\mathcal{M}$. One can define a covariant derivative  $D_a:=X_a^{\mu}\mathcal{D}_{\mu}$  along the coordinates of the worldvolume such that this is also compatible with the background metric. \\\\
Furthermore, if we take the inner product between these tangent vectors we get the components  of the induced metric 
\begin{equation}
g_{ab}=X_a\cdot X_b= G_{\mu\nu}X^{\mu}_aX^{\nu}_b.
\label{induced_metric_comp}
\end{equation}
Here and henceforth a central dot denotes the inner product in the ambient spacetime.
Since $m$ is a hypersurface of $\mathcal{M}$, there 
exists a normal vector, $n$, satisfying the relations
\begin{equation}
n\cdot n =1, \qquad n\cdot X_a=0.
\label{prop. n}
\end{equation}
Therefore, the normal vector and the tangent vectors 
to $m$ form a complete orthonormal basis of the 
spacetime manifold $\mathcal{M}$. In the same spirit, 
the extrinsic curvature, also known as the second fundamental form, is defined as
\begin{equation}
K_{ab}=X_a\cdot D_b n,
\end{equation}
Note that according to (\ref{prop. n}), the extrinsic curvature is symmetric. Both the extrinsic curvature and the induced metric play a fundamental role in the 
modeling of extended objects, as they are used to construct more complex geometric objects~\cite{capovilla1995geometry}. 

To know mechanical properties arising from geometric
models constructed from the fundamental forms, it is
necessary to control the variation of these structures. 
To achieve this, one considers a covariant variation 
denoted as $\cv := \delta X^{\mu} \mathcal{D}_{\mu}$, where $\delta X^{\mu}$ represents the vector along which the deformation occurs. Assuming that the deformation is such that the Lie bracket $[X_a, \delta X]=0$, we have $\cv X_a = D_a \delta X$, then
\begin{equation}
[\cv,D_a]V^{\alpha}=\mathcal{R}^{\alpha}{}_{\beta\mu\nu}V^{\beta}\delta X^{\mu}X_a^{\nu},
\end{equation}
where $V^{\alpha}$ is some vector in $\mathcal{M}$ and $\mathcal{R}^{\alpha}{}_{\beta\mu\nu}$ denotes the 
background Riemann tensor. It is straightforward to
obtain the following variations
\begin{eqnarray}
\cv g_{ab} & = &  2X_{(a}\cdot D_{b)}\delta X,
\\
\cv K_{ab} &= &
- n\cdot D_a D_b\delta X+\mathcal{R}_{\alpha\beta\mu\nu}X^{\alpha}_bn^{\beta}\delta X^{\mu}X_a^{\nu},
\end{eqnarray}
where the round parenthesis stands for symmetrization
in the usual manner. Other important relations are the known Gauss-Weingarten equations
\begin{equation}
\begin{array}{rcl}
 D_a X_b & = & \gamma^c{}_{ab}X_c-K_{ab} n,\\
 D_a n & = & K_{ac}g^{cd}X_d.
\end{array}
\end{equation}
where $\gamma^c{}_{ab}$ are the Christoffel symbols 
of $m$. For a full detail on the geometry of extended objects embedded in curved spaces, 
the reader is referred to
\cite{capovilla1995geometry,capovilla2022covariant}.

\section{\label{sec:citeref} RT model with curved ambient spacetime }
\noindent Considering that the principles governing the evolution of our universe should not differ from the principles that govern the evolution of particles or strings, Regge and Teitelboim proposed that the universe is embedded in a higher-dimensional Minkowski spacetime and its dynamics can be obtained by varying the following action~\cite{regge2016general}
\begin{equation}
S_{RT}=\frac{1}{2\kappa_1} \int_m d^{4}\xi \sqrt{-g} R+ \int_m d^{4}\xi \sqrt{-g} L_m.
\label{flat_action}
\end{equation}
Here, $\kappa_1=8\pi G c^{-4}$  , $R$ is the Ricci scalar of $m$, and  $L_m$ is the matter Lagrangian on the brane.  By varying with respect to $X$, one 
obtains~\cite{regge2016general,karasik2003geodetic}
\begin{equation}
\partial_a\left[\sqrt{-g}\left( G^{ab}-\kappa_1T^{ab}\right)X^{\mu}_a\right]=0,\label{eq flat}
\end{equation}
where $G^{ab}$ is the Einstein tensor and $T^{ab}$ is the energy-momentum tensor of matter fields on the brane. Note that any solution to GR is automatically a solution of (\ref{eq flat}).

If we now allow the spacetime where the brane universe is embedded to be a dynamic 5-dimensional spacetime, the action of the system is given as follows
\begin{equation}
\begin{array}{l}
S=\frac{1}{2\kappa_2}\int_{\mathcal{M}}d^5 y\sqrt{-G}\mathcal{R}+\int_{\mathcal{M}} d^{5}y \sqrt{-G} \mathcal{L}_m+\frac{1}{\kappa_2}\int_m d^{4}\xi \sqrt{-g} K\\
\hspace{1 cm} 
+\frac{1}{2\kappa_1} \int_m d^{4}\xi \sqrt{-g} R+ \int_m d^{4}\xi \sqrt{-g} L_m,
\end{array}
\label{total_action}
\end{equation} 
where $G:= \det (G_{\mu\nu})$, $g:= \det (g_{ab})$, and
$\kappa_2=8\pi G_5 c^{-4}$. Further, $G_5$ is the Newton constant in five dimension while $\mathcal{L}_m$ denotes the matter Lagrangian in the background spacetime, 
and $\mathcal{R}$ is its Ricci scalar. The third element 
in the action corresponds to the Gibbons-Hawking-York (GHY) term, which is crucial to ensure that the variation of the action with respect to the metric of the ambient spacetime is well-defined when boundaries are present. In this context, the brane universe plays the role of a boundary in $\mathcal{M}$ so then this term must be considered. 
We would like to point out that the action~(\ref{total_action}) is close in the spirit to the prototype brane action discussed in \cite{davidson2006dirac}.
Since the ambient space is considered dynamic, varying the action (\ref{total_action}) with respect to the metric $G_{\mu\nu}$ yields the following equations
\begin{equation}
\mathcal{R}^{\mu\nu}-\frac{1}{2}\mathcal{R}G^{\mu\nu}=\kappa_2\left( \mathcal{T}^{\mu\nu}_{bulk}+\mathcal{T}^{\mu\nu}_{brane}\right),
\end{equation}
$\mathcal{T}^{\mu\nu}_{bulk}$ is the energy-momentum tensor of the bulk, and $\mathcal{T}^{\mu\nu}_{brane}$ is the energy-momentum tensor associated with the brane, which is given as follows
\begin{equation}
\mathcal{T}^{\mu\nu}_{brane}=\frac{1}{\sqrt{-G}}\int_md^4\xi\sqrt{-g}\left( T^{ab}-\frac{1}{\kappa_1}G^{ab}\right)X^{\mu}_aX^{\nu}_b\delta^5\left(y^{\mu}-X^{\mu}(\xi)\right).
\label{eq_background}
\end{equation}
Note the particular case where if the equations of general relativity in four dimensions are satisfied, then the brane does not gravitate in the 5-dimensional spacetime, see \cite{cordero2002stealth}.
\\\\
Now, varying (\ref{total_action}) with respect to the embedding functions of the brane universe yields the following dynamic equations
\begin{equation}
D_a\left(\sqrt{-g}\,\mathbf{T}^{ab}X^{\mu}_b\right)- 
\frac{\kappa_1}{\kappa_2} 
\sqrt{-g}\mathcal{R}^{\mu\nu}n_{\nu}=0,
\label{eq_brane}
\end{equation}
where $\mathbf{T}^{ab} := G^{ab} - \kappa_1 T^{ab} - (\kappa_1/\kappa_2) \left( K g^{ab} - K^{ab} \right)$. If the ambient spacetime is flat, the previous equation reduces to the equation $(8)$ of the work \cite{rojas2024dark}.
It is evident from (\ref{eq_brane}) that the curvature of the ambient spacetime plays a fundamental role. This implies that, since the brane is embedded in a curved spacetime, the curvature of this spacetime directly influences the dynamic behavior of the brane. Note also that in this case, when varying with respect to the embedding functions, the GHY term must also be varied, and it directly impacts the dynamic equations of $m$.

\section{\label{cosm} Cosmology in RT model with curved ambient spacetime}

Now we will embed a 4-dimensional FRW spacetime into 
a 5-dimensional curved ambient spacetime and study the resulting brane dynamics. For simplicity, we consider the five-dimensional metric
\begin{equation}
ds_5^2=- \Phi(\tau) d\tau^2+\Psi(\tau) dR^2+\tau^2d\Omega_{3,k},
\label{metric_bg}
\end{equation}
where $d\Omega_{3,k}=\frac{dr^2}{1-k r^2}+r^2 d\theta^2+r^2 \sin^2\theta$ and $\dot{a}=\frac{d a}{dt}$,
is the line element of the three-dimensional spacelike
hypersurface which is assumed to be homogeneous and isotropic.
Following~\cite{akbar2017embedding} we define the embedding functions 
\begin{equation}
X^{\mu}=
\left(\begin{matrix}
a,\\
\int \sqrt{(\Phi\dot{a}^2-1)/\Psi}dt,\\
r,\\
\theta,\\
\phi
\end{matrix}\right),
\end{equation}
and substituting them in (\ref{metric_bg}), we obtain
\begin{equation}
ds_4^2=ds_5^2\big\vert_{m}=-dt^2+a^2d\Omega_{3,k}.
\label{metric_brane}
\end{equation}
Under previous conditions we consider a perfect fluid on the brane, so that the metrics in (\ref{metric_bg}) and (\ref{metric_brane}), along with the embedding functions in equation (\ref{eq_brane}) with $\mu = R$, lead to the following conservation law
\begin{equation}
\fl
\partial_t \left[ a^3 \left( -\kappa_1\rho+\frac{3\left( k+\dot{a}^2\right)}{a^2}+3\frac{\kappa_1}{\kappa_2}\sqrt{\frac{\dot{a}^2}{a^2}-\frac{1}{\Phi a^2}}\right)\sqrt{\Psi\Phi \dot{a}^2-\Psi}+\frac{\kappa_1}{\kappa_2}\frac{a^3}{2\sqrt{\Phi\Psi}}\frac{d\Psi}{da}\right]=0 ,
\label{eq1}
\end{equation}
where $\rho$ is the matter density and $k$ is the spatial curvature. Note that, $\Phi$ and $\Psi$ are functions of the scale factor $a$.

On the other hand, the component $\mu = \tau$ of (\ref{eq_brane}), yields the continuity equation
\begin{equation}
\dot{\rho}+3\frac{\dot{a}}{a}\left(\rho+P\right)=0,
\label{eq2}
\end{equation}
where $P$ represents the pressure of the fluid on the brane. Finally, the other three equations in (\ref{eq_brane}) vanishing identically.
At this point, if we assume for simplicity that $\Phi = 1/\Psi = \lambda^2/a^2$, where $\lambda$ is a constant, and also that $\rho = 0$, $P = 0$, and $k = 0$, then equation (\ref{eq2}) disappears, and equation (\ref{eq1}) can be expressed as follows
\begin{equation}
\left( \lambda^2 H^2+\alpha\sqrt{\lambda^2 H^2-1}\right)\sqrt{\lambda^2 H^2-1}+\frac{\alpha}{3}=\beta\left( 1+z\right)^4,
\label{model}
\end{equation}
where $H$ is the Hubble parameter, $z$ is the redshift, $\alpha = \frac{\lambda \kappa_1}{\kappa_2}$, $\beta = \frac{\omega \lambda^3}{3}$, and $\omega$ is the constant that appears when integrating equation~(\ref{eq1}). 
In fact, within the unified brane cosmology~\cite{karasik2003geodetic}, it parameterizes the 
deviation from the Randall-Sundrum brane cosmology
and from GR when $\omega = 0$.
Despite these simplified conditions, they are sufficient
to test background curvature effects on the dynamics
of the brane. Indeed, by considering~(\ref{model}), the best fit to be in agreement with the actual 
observational data are given in Table \ref{table1} 
for the following values: $\lambda = 0.013 \mathrm{km}{}^{-1}\mathrm{s\ Mpc}$, $\alpha = 0.010$, and $\beta = 0.234$.
\begin{figure}[h]
\centering
\includegraphics[width=0.8\textwidth]{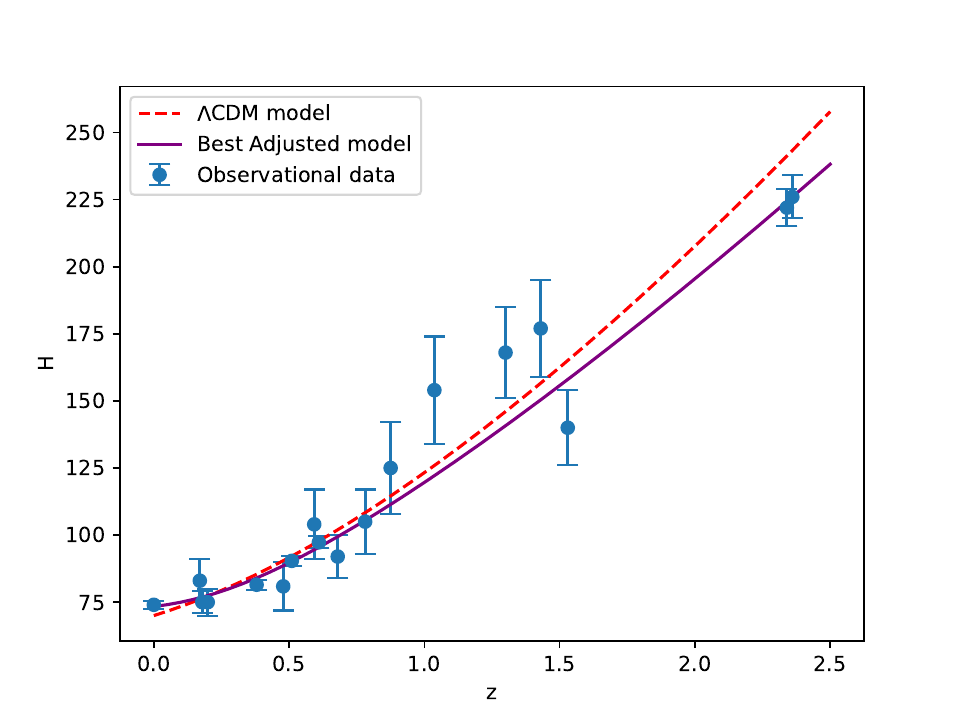}
\caption{\label{fig1} The purple line represents the best fit of the model (\ref{model}), while the red dashed line corresponds to the $\Lambda$CDM model. Here, $H$ is expressed in units of kms${}^{-1}$Mpc${}^{-1}$.}
\end{figure}
\noindent We also created the plot of $H/(1+z)$ versus $z$ to identify the point at which the brane universe transitions from deceleration to acceleration
\begin{figure}[H]
\centering
\includegraphics[width=0.8\textwidth]{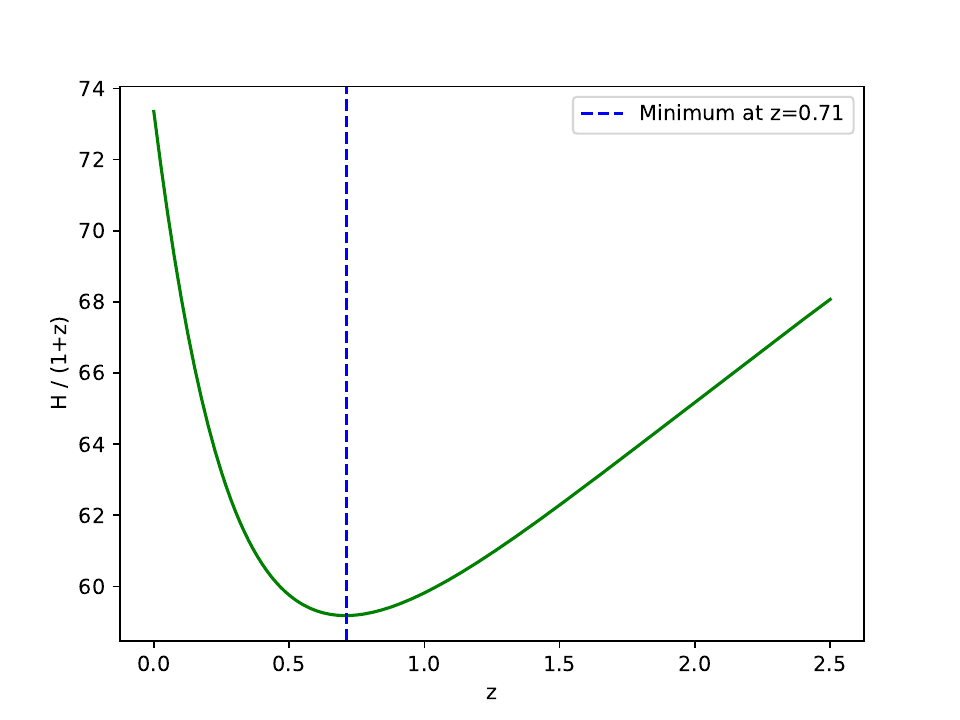}
\caption{\label{fig2} The point $z = 0.71$ corresponds to the transition from a decelerating universe to an accelerating universe. }
\end{figure}
\noindent According to Figure \ref{fig2}, the late-time acceleration begins at $z = 0.71$,
a value within the range predicted by the $\Lambda$CDM model.

The development outlined above changes slightly
if one considers baryonic matter on the brane, for example 
with $\rho \propto a^{-3}$. Indeed,~  Eq.~(\ref{eq1}) can be rewritten as follows
\begin{equation}
\gamma = \frac{\lambda^2 H^2}{(1+z)^3} - \beta \frac{1+z}{\sqrt{\lambda^2 H^2 - 1}} + \alpha \frac{ (\lambda^2 H^2 - 1)^{1/2}}{(1+z)^3} \left[ 1 + \frac{ (\lambda^2 H^2 - 1)^{-1}}{3} \right],
\label{eqm}
\end{equation}
where $\gamma$ is a new parameter associated with the amount of baryonic matter. In this case, the best fit for the model described in (\ref{eqm}) is given by $\alpha = 0.139$, $\beta = 0.259$, $\lambda = 0.014 \ \mathrm{km}{}^{-1}\mathrm{s \ Mpc}$, and $\gamma = 0.01$.
In this way we have a slight improvement in the 
agreement with the predictions provided by the $\Lambda$CDM model.
\begin{figure}[h]
\centering
\includegraphics[width=0.8\textwidth]{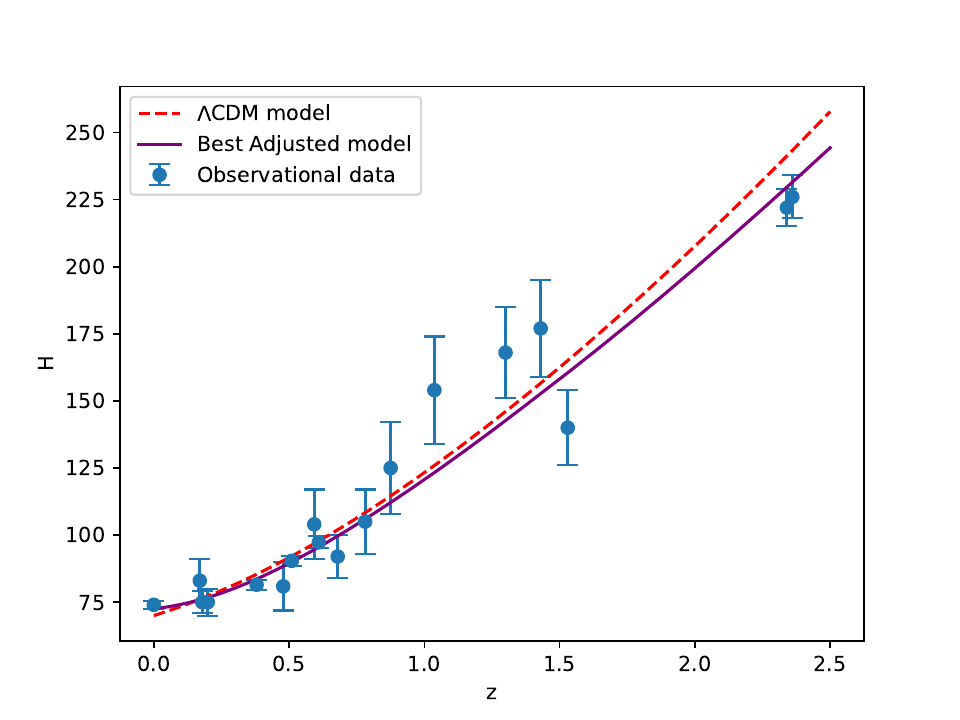}
\caption{\label{fig1} The purple line represents the best fit of the model in (\ref{eqm}).}
\end{figure}
\begin{figure}[h]
\centering
\includegraphics[width=0.8\textwidth]{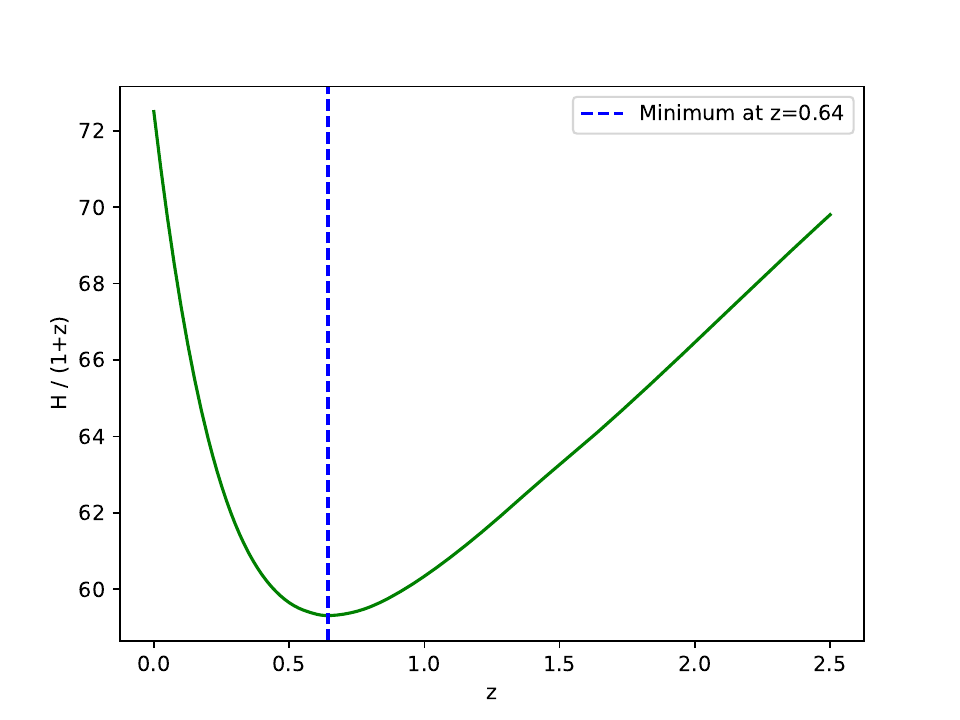}
\caption{\label{fig1}Here, the transition from a decelerating universe to an accelerating universe begins at $z=0.64$, a value consistent with the $\Lambda$CDM model. }
\end{figure}
\begin{table}[H]
\centering
\caption{Data used for fit in Figure \ref{fig1}. Data was selected with $ \sigma_H< 0.15H$.}
\footnotesize
\begin{tabular}{llll}
\br
\textrm{$z$}&\textrm{$H$}&\textrm{$\sigma_H$}&\textrm{Reference}\\
\mr
0 & 74.03 & 1.42 & \cite{riess2019large}\\
0.17 & 83 & 8 & \cite{stern2010cosmic}\\
0.1791 & 75 & 4 & \cite{moresco2012improved}\\
0.1993 & 75 & 5 & \cite{moresco2012improved}\\
0.38 & 81.5 & 1.9 & \cite{alam2017clustering}\\
0.4783 & 80.9 & 9 & \cite{moresco20166}\\
0.51 & 90.4 & 1.9 & \cite{alam2017clustering}\\
0.5929 & 104 & 13 & \cite{moresco2012improved}\\
0.61 & 97.3 & 2.1 & \cite{alam2017clustering}\\
0.6797 & 92 & 8 & \cite{moresco2012improved}\\
0.7812 & 105 & 12 & \cite{moresco2012improved}\\
0.8754 & 125 & 17 & \cite{moresco2012improved}\\
1.037 & 154 & 20 & \cite{moresco2012improved}\\
1.3 & 168 & 17 & \cite{stern2010cosmic}\\
1.43 & 177 & 18 & \cite{stern2010cosmic}\\
1.53 & 140 & 14 & \cite{stern2010cosmic}\\
2.34 & 222 & 7 & \cite{delubac2013baryon}\\
2.36 & 226 & 8 & \cite{font2014quasar}\\
\br
\end{tabular}\\
\label{table1}
\end{table}
\normalsize

\section{\label{conclusion}Conclusions}

In this work, the equations of motion for a brane universe embedded in a dynamic five-dimensional spacetime were discussed. It was subsequently demonstrated that considering an FRW spacetime embedded in a specific curved spacetime within this model can mimic the effects of 
the observed late-time acceleration. Moreover, within this framework, the transition point from a decelerating universe to an accelerating one aligns with the predictions of the $\Lambda$CDM model. Additionally, the agreement improves even further when baryonic matter is included on the brane. This suggests that if our universe is indeed a brane embedded in a higher-dimensional space, several observed effects could be attributed to this embedding. However, caution is in
order. Although this model does not require exotic energy or matter terms on the brane to reproduce these effects, the introduction of additional dimensions entails a significant conceptual cost. In this sense, 
The approach followed here is free of background
matter content, and of a cosmological constant defined on this 
type of universe with the mere intention of highlighting 
the role that the boundary term provided by $K$ plays in the development. In this regard, we believe that it is mainly 
the $K$-term that is responsible for the peculiar acceleration behaviour for this universe at late times. Hence
our model, as an effective modified gravity theory, deserves
more analysis or improvement.
Finally, exploring other physical systems through this approach is important, as it allows for a critical analysis of their implications, physical coherence, and consistency with observational data.

\section*{Acknowledgments}

GC is grateful to Cuauhtemoc Campuzano for discussions and helpful suggestions. ER acknowledges encouragement from ProDePMexico, CA-UV-320: Algebra, Geometría y Gravitación.  GC acknowledges support from a Postdoctoral Fellowship by Estancias Posdoctorales por México 2023(1)-CONAHCYT . Also, ER thanks partial
support from Sistema Nacional de Investigadoras e Investigadores de México.

\section*{References}

\bibliography{apssamp}

\end{document}